\documentclass[prb,twocolumn,showpacs,preprintnumbers,amsmath,amssymb]{revtex4}

\usepackage{graphicx}

\begin{document}

\title{Incommensurate Transverse Anisotropy Induced by Disorder 
and Spin-Orbit-Vibron Coupling in Mn$_{12}$-acetate}
\author{Kyungwha Park$^{1,2}$, Mark R. Pederson$^{1}$, Tunna Baruah$^{3}$, 
Noam Bernstein$^{1}$, Jens Kortus$^{4}$, Steven L. Richardson$^{1,3}$,
Enrique del Barco$^{5}$, Andrew D. Kent$^{5}$, Steve Hill$^{6}$, and Naresh S. Dalal$^{7}$ }
\affiliation{
$^1$Center for Computational Materials Science - 6392, Naval
Research Laboratory, Washington, DC 20375-5000 \\
$^2$Department of Physics, Georgetown University, Washington, DC 20057 \\
$^3$Department of Electrical and Computer Engineering, Howard University, 
Washington DC 20059 \\
$^4$Department of Physics, Max-Plank-Institute for 
Festkerporforschung, Heisenbergstrasse 70569 Stuttgart, Germany \\
$^5$Department of Physics, New York University 4 Washington Place, 
New York, NY 10003  \\
$^6$Department of Physics, University of Florida, Gainesville, FL 32611 \\
$^7$Department of Chemistry and Biochemistry and NHMFL, Florida State University, 
Tallahassee, FL 32306 } 
\date{\today}

\begin{abstract}

It has been shown within density-functional theory that in Mn$_{12}$-acetate 
there are effects due to disorder by solvent molecules and a coupling 
between vibrational and electronic degrees of freedom. 
We calculate the in-plane principal axes of the second-order 
anisotropy caused by the second effect and compare them with 
those of the fourth-order anisotropy due to the first effect. We find 
that the two types of the principal axes are not commensurate
with each other, which results in a complete quenching of the 
tunnel-splitting oscillation as a function of an applied transverse field. 

\end{abstract}

\maketitle

The observation of resonant tunneling of magnetization in the single molecule
magnet Mn$_{12}$-acetate\cite{LIS80} (hereafter Mn$_{12}$) with a ground-state 
spin of $S=10$ has led to many experimental and theoretical investigations.
\cite{FRIE96,BARR97,PEDE99,PARK03,PARK04-Mn12}
A simple anisotropy Hamiltonian for $S=10$ provides an excellent approximation
to the physics occurring at low temperatures. Deriving the Hamiltonian from 
density-functional (DF) calculations is challenging but possible.
\cite{PEDE99,PARK03,PARK04-Mn12} For a particular molecular geometry, 
a magnetic anisotropy tensor 
may be calculated considering spin-orbit coupling within a DF
framework. In the principal-axes coordinates, the lowest-order spin Hamiltonian 
can be simplified to the following form:
\begin{eqnarray}
{\cal H}_0=DS_z^2 + E(S_x^2-S_y^2) \;,
\end{eqnarray}
where $D$ and $E$ are the uniaxial and 
second-order transverse anisotropy parameters and $S_z$ is 
the easy-axis component of the spin operator $S$. The $S_4$ symmetry of 
the ideal Mn$_{12}$ molecule causes the value of $E$ to vanish 
and the lowest-order transverse terms to be fourth order. Only transverse anisotropy
terms are responsible for the resonant tunneling between energy levels that
are almost degenerate. Magnetic tunneling measurements, however, showed that 
some of the resonant tunneling occurred at a level lower than fourth order.\cite{FRIE96} 
To understand this anomaly in the tunneling, Cornia {\it et al.}\cite{CORN02}
proposed that inherent disorder in 
solvent molecules may break the $S_4$ symmetry and provide nonzero
values of $E$. Recent electron paramagnetic resonance (EPR) experiments\cite{HILL03,DELB04}
and magnetic tunneling measurements\cite{DELB03,DELB04} revealed that the model of 
Cornia {\it et al.}\cite{CORN02} needed to be refined for a quantitative comparison 
with experiment. Our DF calculations\cite{PARK04-Mn12} on the Mn$_{12}$ 
showed that the values of $E$ for possible configurations of the disordered solvent 
molecules are in good quantitative agreement with experiment. 

For the Mn$_{12}$, the calculated $D$ value is -0.556~K,\cite{PEDE99}
which agrees well with experiment. The calculated (measured) value of $D$, 
however, does not account for all the measured anisotropy barrier of 
65~K.\cite{BARR97,HILL03} 
One needs a longitudinal anisotropy of order higher than second-order.
In this spirit a fourth-order spin-orbit-vibron (SOV) interaction
was proposed by Pederson {\it et al}.\cite{PEDE02} Ideally, this 
SOV interaction can only contribute to modifying the second-order barrier 
and to activating two fourth-order transverse terms in the spin Hamiltonian. 
At this level a strong SOV interaction can complicate the tunneling
experiments in several different ways. The simplest complication is
that there are two different longitudinal 4th-order terms which 
scale as $S^2S_z^2$ and $S_z^4$ respectively. If the $S_z^4$ term
is dominant, then at any resonant field only a single pair of states
are involved with tunneling.
The fourth-order transverse terms result in departures 
of the period of the tunnel-splitting oscillation from 
$\Delta H_x=2\sqrt{2E(|D|+E)}/g \mu_B$ that has been derived by Garg.\cite{GARG93} 
Such departures were first identified by Wernsdorfer {\it et al.}\cite{WERN99} 
and later modified to include the 4th-order transverse terms by the 
group of Garg.\cite{KECE03}
Recent EPR and magnetic experiments\cite{DELB04} suggested that 
for Mn$_{12}$ the in-plane principal axes of the second-order 
anisotropy may not be aligned with those of the fourth-order anisotropy
so that the oscillation in the tunnel splitting could be completely
quenched in this material, in contrast to Fe$_8$[\onlinecite{WIEG84}].

In this paper, we perform the DF calculations 
on a statistically weighted collection of three different $E$ values
caused by the solvent disorder and on the fourth-order SOV interaction. 
We then combine these two effects to determine whether DF 
theory (DFT) can qualitatively account for a suppression of oscillations in the 
tunnel splittings that occur in transverse tunneling experiments.

Our calculations have been performed within DFT using the NRLMOL 
program\cite{codes}. The NRLMOL calculates an accurate Gaussian-type orbital representation 
for the self-consistent occupied and unoccupied molecular orbitals. 
We use DFT within the generalized-gradient 
approximation\cite{PERD96} to account for the quantum-mechanical 
behavior of the electrons. The geometries of the molecules are fully relaxed. 
Considering the spin-orbit coupling for a relaxed geometry, as discussed
in Ref.~[\onlinecite{PEDE99}], the second-order 
anisotropy Hamiltonian ($\Sigma_{a,b=x,y,z}\gamma_{ab}S_aS_b$) 
may be derived from the calculated orbitals.
Diagonalization of the $\gamma$ matrix provides the 2nd-order anisotropy
barrier and the principal axes.

\begin{figure}[t]
\includegraphics[angle=0,width=0.25\textwidth]{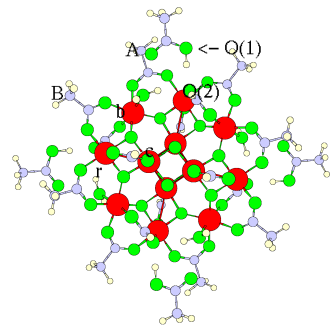}
\caption{Geometry of a Mn$_{12}$ molecule with four hydrogen-bonded acetic
acids molecules. This high-symmetry configuration accounts for
only 1/16 of the total concentration. For details, see Ref.[\onlinecite{PARK04-Mn12}].}
\label{passmol}
\end{figure}

As shown in Fig.~\ref{passmol}, a single Mn$_{12}$ molecule 
is surrounded by four acetic-acid solvent molecules (CH$_3$CO$_2$H), 
each of which is shared by two neighboring Mn$_{12}$ molecules. 
If a Mn$_{12}$ molecule is hydrogen-bonded to four CH$_3$CO$_2$H
molecules as pictured in Fig.~\ref{passmol}, then
the symmetry of this molecule remains the same as $S_4$. 
Since CH$_3$CO$_2$H can bind to either of the Mn$_{12}$
neighbors with the same energy, 1/16 of the Mn$_{12}$ molecules 
have the S$_4$-symmetry with hydrogen bonds and another 1/16 of the molecules 
have the S$_4$ symmetry {\it without} hydrogen bonds. The remaining molecules 
have different orientations which break the symmetry. As discussed in 
Ref.~\onlinecite{PARK04-Mn12},
there are a total of six different configurations which have the number of
hydrogen bonds $n$, statistical weights, and $D$ and $E$ values of 
$(0,1/16,-0.54,0.000)$, $(4,1/16,-0.56,0.000)$, $(1,4/16,-0.54,0.008)$, 
$(2,4/16,-0.55,0.000)$, $(2,2/16,-0.55,0.016)$, and $(3,4/16,-0.55,0.008)$. 
The two distinctive cases of two hydrogen bondings are characterized by having 
neighboring acetic acids ("cis") hydrogen-bonded or next neighboring acetic acids 
("trans") bonded, respectively. The S$_4$ symmetry suggests that in the case of 
significantly small off-diagonal elements in the $\gamma$ matrix for $n=1$, 
the perturbations due to the "cis"-geometry should cancel each another and 
that those due to the "trans"-geometry should add constructively. Thus, we 
find that 5/8 of the molecules have appreciable $E$ values of 0.008-0.016~K.

The Hamiltonian for a single uniaxial anisotropic spin coupled to a 
one-dimensional harmonic oscillator is given by 
${\cal H}= \gamma_{zz} S_z^2 + \frac{1}{2}(P^2+\omega^2 Q^2) + Q
\sum_{ab} \gamma_{ab}^{'} S_aS_b$, where $P$, $Q$, and $\omega$ are 
the momentum, position, and frequency of the oscillator and 
$\gamma_{ab}^{'}=d\gamma_{ab}/dQ$. As shown in Ref.~[\onlinecite{PEDE02}], 
the energy of the coupled system as a function of an eigenvalue 
of $S_z$ ($M$) is 
${\cal E}= \omega /2 + D M^2 - (A+BM^2)^2/2\omega^2$
where $A= S(S+1)(\gamma_{xx}^{'}+\gamma_{yy}^{'})/2$
and $B= \gamma_{zz}^{'}-(\gamma_{xx}^{'}+\gamma_{yy}^{'})/2$.  
Generalizing this problem to the Mn$_{12}$
requires coupling all normal modes. 
In general, the fourth-order corrections to the Hamiltonian are 
${\cal H}_4= \Sigma_{abcd} A_{abcd} S_a S_b S_c S_d$.         
Only the Raman active vibrational modes lead to a change
in the fourth-order barrier. 

Combining the SOV interaction with the symmetry-breaking effects of 
the solvent disorder, we can write the total spin Hamiltonian as:
\begin{eqnarray} 
{\cal H} &= & DS_z^2 + E[\cos(2\alpha)(S_x^2-S_y^2)+2\sin(2\alpha)S_xS_y] 
\nonumber \\
&&+ A_1S_z^2(S_z^2-S^2/3) + A_2[3S^4+35S_z^4-30S^2S_z^2] \nonumber \\
&&+ B_1 (S_x^4+S_y^4-6S_x^2S_y^2) + B_2 [S_xS_y(S_x^2-S_y^2)] \:,
\label{eq:totham}
\end{eqnarray} 
where $\alpha$ denotes the angle between the medium axes of the calculated 
second-order and fourth-order anisotropy and $A_1$, $A_2$, $B_1$ and $B_2$
represent fourth-order anisotropy parameters. 
If one replaces the operators $(S_x,S_y,S_z)$ by classical spin vectors and
recast the fourth-order transverse terms in spherical coordinates, 
then the classical potential energy in the $xy$ plane 
(e.g. $\theta=\pi/2$) becomes 
\begin{equation} 
{\cal E} =  {\cal E}_0 + S^4[B_1 \cos(4\phi) + \frac{B_2}{4} \sin(4\phi)] \:,
\label{eq:E4}
\end{equation} 
where ${\cal E}_0$ is a constant term and $\phi$ is the azimuthal angle.
This shows that the energy surface of a system with S$_4$ symmetry,
at fourth order, resembles a four-leaf clover in the $xy$ plane. Furthermore, when
the fourth-order terms in Eq.~(\ref{eq:E4}) are accounted for, the
in-plane principal axes are determined to lie along the nodes or antinodes
of the classical potential. In addition, once $B_1$ and $B_2$ are determined 
in given coordinates, a rotation by 
$\cos^{-1}[B_1/(B_1^2+B_2^2/16)^{1/2}]$ allows one to  
re-express the two transverse terms as $C(S_+^4+S_-^4)$ with 
$C=[B_1^2+B_2^2/16]^{1/2}/2$. The experimental value of $C$  
is $2.3 \times 10^{-5}$K,\cite{HILL03} while the SOV contributions 
to $B_1$ and $B_2$ lead to C$_{\mathrm{sov}}=0.053 \times 10^{-5}$K.\cite{PEDE02} 

\begin{figure}
\includegraphics[angle=0,height=0.25\textwidth,width=0.48\textwidth]{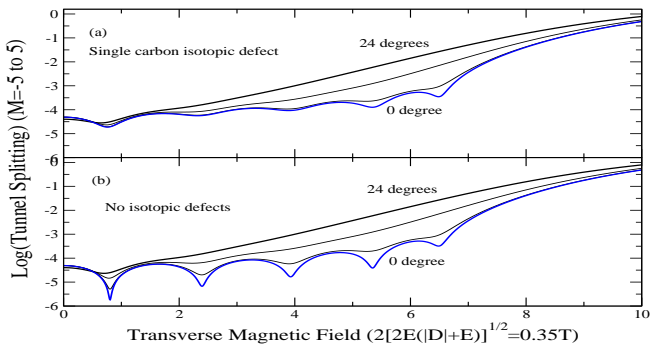}
\caption{Logarithm of the $M=\pm 5$ tunnel splitting versus applied transverse 
field for a Mn$_{12}$ with one $^{13}$C isotopic defect and pure Mn$_{12}$. Different 
values for $\alpha$ (0, 3, 12, and 24$^{\circ}$) were used.}
\label{fig:tunnel}
\end{figure}

Even with the rotation of the $B_1$ and $B_2$ terms to the more 
usual definition, the 2nd-order transverse terms may not, in general,
be simultaneously expressed in the diagonal form. So the dependence on
$\alpha$ is required. To compute $\alpha$ we calculate the separate
magnetic anisotropy energies from the SOV interaction and the solvent 
disorder as a function of $\phi$ in the $xy$ plane. For the solvent
disorder only three configurations [$n=1$, 2(trans),and 3] are considered
in the calculation because the rest of the three configurations do not
produce $E$. We find that $\alpha$ is 24$^{\circ}$. This is a bit 
smaller than the experimentally measured value of $\alpha=27^{\circ}$.
\cite{DELB04} We calculate the tunnel splitting between $M=-5$ and
$M=5$ with various values of $\alpha$ in zero field from diagonalizing 
the quantum mechanical form of Eq.~(\ref{eq:totham}) 
which takes into account the non-commutativity of the spin operators. 
For the pure Mn$_{12}$ sample we find that the oscillation in the
tunnel splitting is entirely quenched at the calculated value of $\alpha=24^{\circ}$
as shown in Fig.~\ref{fig:tunnel}(b). When $\alpha=0$, as expected from 
Fe$_8$,\cite{WIEG84} the oscillation is 
clearly visible. Even for a very small value of $\alpha$ such as 3$^{\circ}$, 
the amplitude of the oscillation tends to be mostly damped out. As the applied 
transverse field increases, the effect of the incommensurate transverse anisotropy 
decreases because the tunneling is governed by a strong transverse field. 
Because of the fourth-order terms in Eq.~(\ref{eq:totham}), the period of the 
oscillation decreases as the transverse field increases.\cite{KECE03} 

For a naturally occurring collection of Mn$_{12}$ molecules, 
37\% of the molecules contain one or two isotopic defects. In the case of
isotopic defects in the Mn$_{12}$, the SOV interaction provides 
additional symmetry-breaking in the spin Hamiltonian. If the SOV interaction 
is large enough, this fraction of molecules would contribute to 
an almost continuous distribution of broken-symmetry anisotropy 
Hamiltonians. We calculate the same tunnel splitting for the Mn$_{12}$ 
doped with one carbon isotope ($^{13}$C) for various values of $\alpha$
[Fig.~\ref{fig:tunnel}(a)]. We find that even 
one isotopic defect substantially suppresses the oscillation in the tunnel 
splitting but the period remains unperturbed.
Since the tunneling experiments were performed in the presence of the 
longitudinal field that makes $M$ and $M^{\prime}$ almost degenerate 
where $M+M' \neq 0$, direct comparison with experiment is not possible. 
The effect of the longitudinal field and the different types of isotopic 
defects on the tunnel-splitting is in progress.

In summary, for Mn$_{12}$ we have considered the second-order 
anisotropy induced by the disordered solvent molecules
and the fourth-order SOV interaction within the DFT framework. 
The second-order transverse terms could cause the oscillation in the tunnel 
splitting. Experimental data indicated that the in-plane principal axes of 
the second-order anisotropy was not commensurate with those of the fourth-order 
anisotropy. DFT calculations showed that the angle between
the two types of anisotropy and resulting quenching of the
oscillation agree well with experiment.

This work was supported in part by ONR N000140211045, NSF Grant Nos. HRD-0317607, 
DMR-0103290, DMR-0239481, DMR-0114142, DMR-0315609, and the DoD HPCMO CHSSI program. 
A.D.K. and M.R.P. acknowledge the hospitality of the Aspen Center for Physics where this
collaboration was initiated.

\clearpage
\end{document}